\documentclass[useAMS,usenatbib]{mn2e}
\pdfoutput=1

\usepackage{url}
\usepackage{color}
\usepackage{lscape}
\usepackage{supertabular}
\usepackage{graphicx}
\usepackage{chngpage}
\usepackage{booktabs}

\newcommand{\milliJans}{\,{\rm mJy}}

\newcommand{\mjPerArcsec}{\milliJans\,{\rm s^{-1/2}}}

\title[Automated rapid follow-up of \textit{Swift} GRBs with AMI-LA]
  {Automated rapid follow-up of \textit{Swift} GRB alerts at 15\,GHz with the AMI Large Array}
\author[T. Staley et al.]
  {T.D.~Staley,$^1$
  D.J.~Titterington,$^2$
  R.P.~Fender,$^1$
  J.D.~Swinbank,$^3$
  A.J.~van~der~Horst,$^3$
  \newauthor 
  A.~Rowlinson,$^3$ 
  A.M.M.~Scaife,$^1$
  K.J.B.~Grainge,$^{2,4}$
  G.G.~Pooley$^2$  
  \\
  $^1$School of Physics \& Astronomy, University of Southampton, Southampton SO17 1BJ\\
  $^2$Astrophysics Group, Cavendish Laboratory, 19 J.~J.~Thomson Avenue, Cambridge CB3 0HE\\
  $^3$Astronomical Institute Anton Pannekoek, Science Park 904, P.O. Box 94249, 1090 GE Amsterdam\\
  $^{4}$Kavli Institute for Cosmology Cambridge, Madingley Road, Cambridge CB3 0HA \\
  }
\begin{document}

\date{In original form 2012 September 28}

\pagerange{\pageref{firstpage}--\pageref{lastpage}} \pubyear{2002}

\maketitle

\label{firstpage}

\begin{abstract}
We present 15-GHz follow-up radio observations of eleven \textit{Swift} gamma-ray burst (GRB) sources, obtained with the Arcminute Microkelvin Imager Large Array (AMI-LA). 
The initial follow-up observation for each source was made in a fully automated fashion; as a result four observations were initiated within five minutes of the GRB alert timestamp. 
These observations provide the first millijansky-level constraints on prolonged radio emission from GRBs within the first hour post-burst.
While no radio emission within the first six hours after the GRB is detected in this preliminary analysis, radio afterglow is detected from one of the GRBs (GRB120326A) on a timescale of days. 
The observations were made as part of an ongoing programme to use AMI-LA as a systematic follow-up tool for transients at radio frequencies. 
In addition to the preliminary results, we explain how we have created an easily extensible automated follow-up system, describing new software tools developed for astronomical transient alert distribution, automatic requesting of target-of-opportunity observations, 
and robotic control of the observatory. 
\end{abstract}


\section{Introduction}
\label{sec:intro}
Since the discovery of GRB afterglow emission at X-ray \citep{Costa1997}, optical \citep{Paradijs1997} and radio frequencies
\citep{Frail1997}, it has been demonstrated that broadband observations across the electromagnetic spectrum can be utilized to determine the micro- and macro-physical parameters of the GRB explosion and its immediate environment \citep[e.g.][]{Sari1998,Wijers1999}. 
After the initial detection in gamma- and X-rays, rapid follow-up observations are often essential to capture the GRB afterglow in the optical, before it fades. 
To this end, automated, robotic optical observations from smaller telescopes are now routine \cite[e.g.][]{Akerlof2003,Wozniak2005}. In contrast, early follow-up observations of GRBs at radio wavelengths are rare  \cite[although some attempts have been made, see e.g.][]{Green1995,Koranyi1995,Dessenne1996}.

In this paper we explain how we have achieved fully automated rapid follow-up observations of GRB events with the Arcminute Microkelvin Imager Large Array (AMI-LA), a 15-GHz aperture synthesis radio telescope located near Cambridge, UK. A primary beam width of 5.5 arcminutes and a sensitivity of $\sim3\mjPerArcsec$ make this an ideal facility for follow-up of the well localised GRB events observed by the \textit{Swift} burst alert telescope \citep[Swift-BAT,][]{Gehrels2004,Barthelmy2005}.

This paper is structured as follows:
First we give the scientific motivation for obtaining early-time radio follow-up observations (Section~\ref{sec:science}). 
A brief review of the relevant models for GRB emission is given, 
focusing upon predictions for radio emission at the time of the gamma ray emission and within the first few hours thereafter. 
We then explain how this programme will provide novel data, complementary to existing GRB follow-up catalogues.
In Sections~\ref{sec:ami} -- ~\ref{sec:data} we describe the AMI Large Array; the system we have put in place to initiate automated follow-up; and our data reduction techniques.
In Section~\ref{sec:results}, we report on system response times and present 15-GHz lightcurves. 
Finally, we describe our plans for further analysis of the AMI-LA data, and possibilities for follow-up with alternative facilities, in Section \ref{sec:discuss}.

\section{Scientific motivation}
\label{sec:science}
\subsection{GRB radio emission at short and long timescales}
The `relativistic fireball' is the prevalent model to describe GRB emission across the electromagnetic spectrum. In this model a blast wave propelled by an ultra-relativistic jet ploughs through an external medium. Electrons accelerated by the shock emit synchrotron radiation in the magnetic field, which is also amplified by the shock \citep{Rees1992,Meszaros1997}. 
Radio observations are crucial for pinning down the evolution of the peak flux and two of the three characteristic frequencies of the broadband synchrotron spectrum: the peak frequency and the synchrotron self-absorption frequency (with the third one being the cooling-break frequency). 
The radio emission from the blast wave typically peaks much later than at optical and X-ray wavelengths, on a time-scale of days to weeks depending on the observing frequency \citep{Chandra2012}, and in some cases can be observed months or even years after the initial GRB explosion \citep[e.g.][]{Frail2000b,Horst2008}. 
Although radio afterglows are quite faint, with the majority having sub-millijansky peak fluxes \citep{Chandra2012}, a large number of these have well-sampled lightcurves at timescales from days to years (see Figure~\ref{fig:cf2012}).

There are a few GRBs for which the radio observations started within the first day after the gamma-ray trigger, and some resulted in the detection of an early peak \citep[e.g.][]{Kulkarni1999,Frail2000a}. The favoured explanation for these early radio flares is emission from the `reverse shock,' a phenomenon arising from dynamics at the shock front. As the shell of shocked matter slows and the surrounding medium begins to influence the dynamics, the interaction region becomes bounded by two shock fronts: one propagating into the unshocked external medium (`forward shock'), and a second, `reverse shock' propagating backwards into the relativistic flow \citep[as seen from the flow rest frame; see e.g.][]{Kobayashi1999}. 
This model may be complicated further if there are pre-existing shock fronts in the external medium due to stellar winds \citep[][]{Peer2006}. 
Emission from the reverse shock region may give rise to high flux levels across the frequency spectrum at much earlier times than the typical afterglow, and it has been predicted that radio flares due to the reverse shock are more likely to occur within the first hours after the GRB \citep{Melandri2010}. 

Although a reverse shock should in principle be formed, the brightness may vary considerably, perhaps resulting in only a small fraction of GRBs displaying radio flares. However, this phenomenon is poorly constrained due to the scarcity of radio observations soon after the prompt gamma-ray emission.
With the programme at AMI discussed in this paper we will 
be able to systematically probe this part of observational parameter space and look for radio flares caused by reverse shock emission. 
The detections of, or upper limits on, early peaks at radio frequencies are important for putting constraints on the physical parameters of the blast wave, for instance its energy, the energy in electrons, and the magnetic field.  
Furthermore, they can help solve an outstanding issue in GRB jet formation, namely whether the jet is Poynting flux dominated or a baryonic outflow, by constraining the energy emitted as the relativistic flow begins to interact with the external medium \citep{Piran2000}. 
This would be most effective in combination with observations constraining very early optical flashes, which have been searched for systematically for some years now \citep[see][and references therein]{Roming2006}. 

Besides the forward shock and reverse shock in GRB jets, there have been predictions of alternative sources of radio emission in the case of short GRBs \citep[those with prompt gamma-ray emission durations of less than $\sim$2 seconds,][]{Kouveliotou1993}. 
One favoured model for short GRBs is the merger of a binary system consisting of two neutron stars or a neutron star and a black hole. 
These mergers are thought to produce mildly to sub-relativistic outflows, resulting in radio emission on timescales of weeks to years \citep{Nakar2011,Piran2012}. On shorter timescales the predictions are more uncertain, one possible outcome being the generation of pulsar-like coherent emission by the rapidly rotating magnetar that is produced shortly after the binary merger \citep{Pshirkov2010}. 

There may even be unpredicted mechanisms that produce early radio emission, and this follow-up programme will provide 
a systematic search for these at $\sim15$\,GHz frequencies.
Intriguingly, recent observations at 1.4\,GHz  made with a 12\,m single dish antenna at the Parkes radio observatory have suggested the presence of short coherent bursts of radio emission within the first 20 minutes after the GRB onset, hinting at the presence of just such radio phenomena occurring at early times \citep{Bannister2012}. 

\subsection{Sample sizes and selection bias}
The sample of GRB radio frequency observations presented in \cite{Chandra2012} --- albeit large --- is far from complete, being both strongly sensitivity limited and biased due to target selection effects.
The observations that have been made, and are still being carried out, 
at different facilities (e.g. Jansky Very Large Array (JVLA), Westerbork Synthesis Radio Telescope, Australia Telescope Compact Array, Giant Metrewave Radio Telescope) are restricted to a subset of possible GRB triggers due to limited available observing time. 
Subsequently only a small fraction of GRBs have been followed up, mostly those that have already shown interesting behaviour in the optical or X-ray bands. 
The same constraints on observing time also restrict systematic follow-up of GRB non-detections, with a radio non-detection in the first few days after the burst often resulting in no further radio follow-up, despite the fact that the peak flux of the radio emission will often occur at later times. 

The recent upgrade of the JVLA allows for much deeper observations, which will help significantly in probing GRB radio afterglows at fainter flux levels.
However, the number of GRBs that the JVLA and other facilities will observe is still limited, as illustrated by the fact that of the eleven GRBs observed and reported upon here, only five have been followed up by other radio telescopes (and the observations made known to the follow-up community). 
Although AMI is not as sensitive as the JVLA, this programme will provide a uniformly selected sample which can be used for comparison with GRB follow-up at other wavelengths. With systematic follow-up we should also detect many 15-GHz radio-afterglow peaks at later times when other observatories have stopped observing.

In summary, after the first year(s) of this project, we will have a uniquely large sample of early time radio observations of GRBs, providing important new constraints on short (minutes to hours) timescale radio emission. We will also provide a systematically obtained and rapidly reduced dataset, flagging up interesting GRBs for deeper observation both with AMI and other observatories as appropriate.

\section{The Arcminute Microkelvin Imager Large Array} 
\label{sec:ami}
The Arcminute Microkelvin Imager Large Array \citep[][]{Zwart2008a} is a synthesis telescope composed of eight equatorially-mounted 12.7\,m dishes sited at the Mullard Radio Astronomy Observatory at Lord's Bridge, Cambridge. 
The telescope observes in the band 12.0--17.9\,GHz with eight 0.75\,GHz   bandwidth channels.
In practice, the two lowest frequency channels   are not generally used due to a lower response in this frequency range
and interference from geostationary satellites, which leaves an effective bandwidth of 13.5--17.9\,GHz and a central frequency of 15.75\,GHz using channels 3--8.
  After calibration, the phase is generally stable to 5$^{\circ}$ for channels 4--7 and 10$^{\circ}$ for channels 3 and 8.
The telescope measures a single linear polarisation (I + Q) and has a flux density sensitivity of $\simeq3\mjPerArcsec$. Flux density calibration follows the Perley-Butler 2010 scale (R. Perley, private communication).  
AMI-LA data reduction has been described extensively in previous works \citep[see e.g.][]{Zwart2008a} and the flux calibration is accurate to better than 5 per cent \citep{Scaife2008,Hurley-Walker2009}. 
During normal operation, the AMI-LA performs continuous execution of
observations scheduled as entries in a queue, with observing programmes
often set up for several days in advance.  Implementation of rapid response
observing for this GRB study, as part of a new target-of-opportunity
capability for AMI, involves automatic insertion of observing requests
into the queue, if necessary displacing existing entries or an observation
in progress.

\section{Automated response system and observing policy}
\label{sec:system}
The problem of automated transient detection and follow-up is multifaceted, with potential for detailed focus in many subfields such as detection, classification, resource allocation, etc. For the purposes of this paper we restrict our discussion to three broadly defined sub-systems: the distribution network, observation request triggering, and the automated telescope response.

\subsection{Transient alerts distribution via the VOEvent network} 
NASA uses the Gamma-ray Coordinates Network \citep[GCN,][]{Barthelmy1998} to rapidly distribute information about GRBs from satellites such as {\it INTEGRAL}, {\it Swift} and {\it Fermi}. Until recently, the most common medium was a short piece of plain text conforming to a specified format, distributed via a dedicated central server / client network. 
While fast and effective for this purpose, this approach will not encompass the range of hierarchical data structures 
that are envisioned for describing the wealth of transient events detected by a new generation of astronomical facilities such as LOFAR, Pan-STARRS, GAIA, LSST etc. 
A robust network of many facilities will also require a more flexible and adaptable distribution model, i.e. one that is largely decentralized. 

The VOEvent standard \citep[see footnote\footnote{`Sky Event Reporting Metadata
Version 2.0,' Rob Seaman et al, 2011: \url{http://www.ivoa.net/Documents/VOEvent/index.html}},][]{Williams2006} has been developed with these needs in mind, providing a flexible data format specified as an XML schema. 
The standard is transport independent (allowing for multiple distribution methods), allows for a distributed network of clients and servers, and has been previously employed by projects such as \mbox{eStar}, the Catalina Real Time Survey, and \mbox{SkyAlert} \citep{Allan2006,Drake2009,Williams2009}. GCN has now started distributing VOEvents (S. D. Barthelmy , private communication) using the VOEvent transport protocol%
\footnote{`VOEvent Transport Protocol,'  A. Allan and R.B. Denny: \url{http://www.ivoa.net/Documents/Notes/VOEventTransport/}},%
which is an interoperable standard for VOEvent distribution.

We tested two freely available software packages implementing the VOEvent transfer protocol: `Dakota'%
\footnote{`Dakota,' R.B. Denny: \url{http://voevent.dc3.com}}%
  and `Comet'%
\footnote{`Comet,' J.D.~Swinbank: \url{http://comet.transientskp.org}}.
Both provide means of generating, distributing and subscribing to streams of VOEvents. 
For this work, we use Comet, primarily because it is implemented in the Python programming language. 
The packages are dedicated to efficient and robust networking; processing of the information enclosed in the packets is delegated to an external program of the user's choice.
In the current system implementation we monitor for new GRB events via a direct connection to the GCN VOEvents service, and pipe received VOEvents to a custom software package described below. 

\subsection{VOEvent monitoring and trigger generation}
The VOEvent identification and response routines were implemented using a Python package written to facilitate such tasks: `pysovo,' which is available under an open-source licence from \mbox{\url{https://github.com/timstaley/pysovo}}.
The package makes it easy for the user to extract information from the VOEvent XML packets; implement decision logic to determine whether an observation request will be triggered; determine whether the source is currently observable from a given observatory; generate requests formatted to custom observatory specifications, and send notification emails and SMS (i.e. text to mobile phone) alerts to designated recipients.
The package is designed to be easily extensible, in order to allow interaction with multiple sources of transient astronomical alerts and multiple follow-up facilities. 

For this programme, we perform VOEvent monitoring and trigger generation using instances of Comet and pysovo running on a dedicated machine in Southampton. Each VOEvent has an ID string, referred to as an `International Virtual Observatory Resource Name,' (IVORN),  which uniquely identifies a VOEvent in much the same way that a URL identifies a resource on the internet.
We use a simple filter based on the IVORN to identify the VOEvent information packets pertaining to new \textit{Swift} GRB alerts. Once relevant VOEvents have been identified, the coordinates are extracted, checked against our observation policy, and used to generate a specially formatted observation request email which is then sent to an address at the Mullard Radio Astronomy Observatory, Cambridge. 

\subsection{Automation of the AMI telescope}
At the telescope level, incoming request emails to a dedicated account 
are picked out by an email filter based on the subject line and a template contained in the message body.
The request email is then passed to a custom program written to respond to AMI target of opportunity requests, `rqcheck,' running on the AMI-LA control computer. 
The email is also forwarded to observatory members, who can act as request moderators  if necessary.

The `rqcheck' 
process validates the email, checking syntax and keyword values, and then proceeds to:
\begin{itemize}
 \item Check whether target is currently on-sky and not obscured by the Sun or Moon.  Sets start and stop times for the
   next possible timing within the next 24 hrs, according to the
   timing specified in the request (ASAP, next transit, or specified sidereal time).
  \item Find the nearest phase calibrator from VLBA\footnote{NRAO VLBA catalogue: \url{http://www.vlba.nrao.edu/astro/calib/}} and JVAS \citep{Patnaik1992} catalogues.
  \item Check the availability of the telescope, e.g. in case of priority observing or engineering work.
  \item Construct an entry for the observing queue.  This will always be a
   standard pointing observation with interleaved calibrator, using the
   default integration times for either array.
  \item Insert this entry in the queue, displacing any previously scheduled
    observations if necessary.
 \item Append the request to a log file.
 \item Notify the requester of the outcome by reply e-mail.
\end{itemize}

\subsection{Trigger and follow-up policy}
For observations with AMI we employ a naive trigger policy, whereby all \textit{Swift} GRB alerts with a declination above -10$^{\circ}$ cause an observation request. 
This simple strategy is optimal in this case as we have access to sufficient observing time to allow systematic follow-up of all \textit{Swift} GRB triggers in the northern hemisphere.
We are employing a logarithmic follow-up schedule for all triggers, observing promptly   after the initial burst, after 1-2 days, 3-4 days, 1 week, 2 weeks,   and finally at 1 month.
Follow up of detected sources will continue to be evaluated manually.
Our standard integration time for each observation is currently one hour.

\section{Data reduction and analysis}
\label{sec:data}
When undertaking systematic observations such as those entailed by a GRB follow-up programme, it is desirable to have a fully automated reduction and analysis pipeline. With this goal in mind we have developed a procedure requiring minimal human invervention, which we plan to fully automate in the near future.
\subsection{Data transfer and image synthesis}
After observations have been made, the raw data are transferred from Cambridge to Southampton for further analysis. 
The data are then reduced using a procedure scripted in Python which performs the following tasks:
\begin{itemize}
\item Identifies set of observations pertaining to each GRB.
\item Runs the AMI software tool `reduce' on each observation to perform
   flagging for interference, shadowing and hardware errors, apply gain
   calibrations and synthesise the frequency channels to produce calibrated
   visibility data in {\it uv}-{\tt FITS} format.
\item Extracts certain metadata which can be used to flag observations
  of poor quality (e.g. system temperature estimate, RFI flagging percentages).
\item Generates images using the casapy `clean' algorithm, including a `dirty' map and PSF map, for visual inspection.
\end{itemize}

\subsection{Image time-series analysis}
Once a time-series of images has been synthesised these are visually checked for for data quality, artefacts, and sources coincident with the known GRB position.
We then attempt to extract a lightcurve by running custom source-fitting software developed for LOFAR data analysis \citep{Spreeuw2010}. 
This allows for source fitting at a known location, in addition to `blind' source finding and extraction above a given threshold. 
To ensure that marginal detections are not missed, we use the source-fitting software to attempt an extraction at the location given by the enhanced \textit{Swift} X-ray Telescope reduction \citep{Burrows2005}; except where this is unavailable (e.g. in the case of GRB120403A) and we resort to the \textit{Swift}-BAT refined position.
In the case of a detection, the code attempts to fit an elliptical Gaussian profile, and errors on the fit are calculated according to the formulae of \cite{Condon1997}. 
When no detection is made, we assign an upper limit of three times the image RMS level.

\section{Results}
\label{sec:results}
\subsection{Response times}
\begin{table}
%
%
%
%
\begin{tabular}{ l  r  r  }
\toprule
 \parbox{ 0.1 \textwidth}{ \begin{center}\textbf{ GRB ID } \end{center} } 
 &   \parbox{ 0.1 \textwidth}{ \begin{center}\textbf{ Hours since burst } \end{center} } 
 &   \parbox{ 0.15 \textwidth}{ \begin{center}\textbf{ 3$\sigma$ upper limit (mJy)  } \end{center} } 
  \\
\toprule

    \textbf{ GRB120305A } &  0.07 &  0.316 \\    

    \textbf{ GRB120308A } &  0.08 &  0.164 \\    

    \textbf{ GRB120311A } &  0.07 &  0.230 \\    

    \textbf{ GRB120320A } &  14.32 &  0.196 \\    

    \textbf{ GRB120324A } &  0.07 &  0.218 \\    

    \textbf{ GRB120326A } &  7.37 &  0.430 \\    

    \textbf{ GRB120403A } &  7.48 &  0.238 \\    

    \textbf{ GRB120404A } &  11.32 &  0.223 \\    

    \textbf{ GRB120422A } &  6.41 &  0.616 \\    

    \textbf{ GRB120514A } &  50.91 &  0.211 \\    

    \textbf{ GRB120521C } &  0.24 &  0.302 \\    

\bottomrule
\end{tabular}

\caption{
Delay between GRB trigger timestamp and the start time of our initial observation, for each GRB followed up.
No radio sources at the position of the GRB were detected at the 3$\sigma$ level in our preliminary analysis of these data, and we list corresponding upper limits for the flux density averaged over the one hour observation period. Note that GRB 120326A has a \textit{marginal} detection at $0.337\milliJans$ in the initial observation, $\sim 7$ hours after the trigger timestamp. 
\label{tab:init_obs}
}
\end{table}
We can estimate system response times by inspecting timestamps on the GCN notices and the AMI observation request logs. 
The typical response time for a target which is immediately observable with AMI is approximately 5 minutes from \textit{Swift}-BAT trigger time to AMI taking data, and we can break this down into contributing factors. 
First, \textit{Swift} GRB positions are broadcast from GCN, with a typical time delay of 7--30 seconds after the trigger time (approximately 0.1 seconds of which is due to GCN --- S. Barthelmy, private communication). 
These are received by our VOEvent node in Southampton, processed, and an email is sent to the AMI control system. 
Round trip time from NASA-GCN to AMI is consistently 4--6 seconds, including the time required to authenticate with the email server. 
Alerts distributed solely via VOEvent transfer protocol  will be relayed with an even shorter delay.
If the target is on the sky, AMI then begins to slew almost immediately.  
The maximum slewing speed of the telescope is approximately $15^{\circ}$ per minute, which means that response
times for immediate observation will be of the order of a few minutes; currently a 4-minute period is allotted to ensure that the telescope is on source before taking data. 

\begin{figure}
\begin{center}  
    \includegraphics[width=0.48\textwidth]{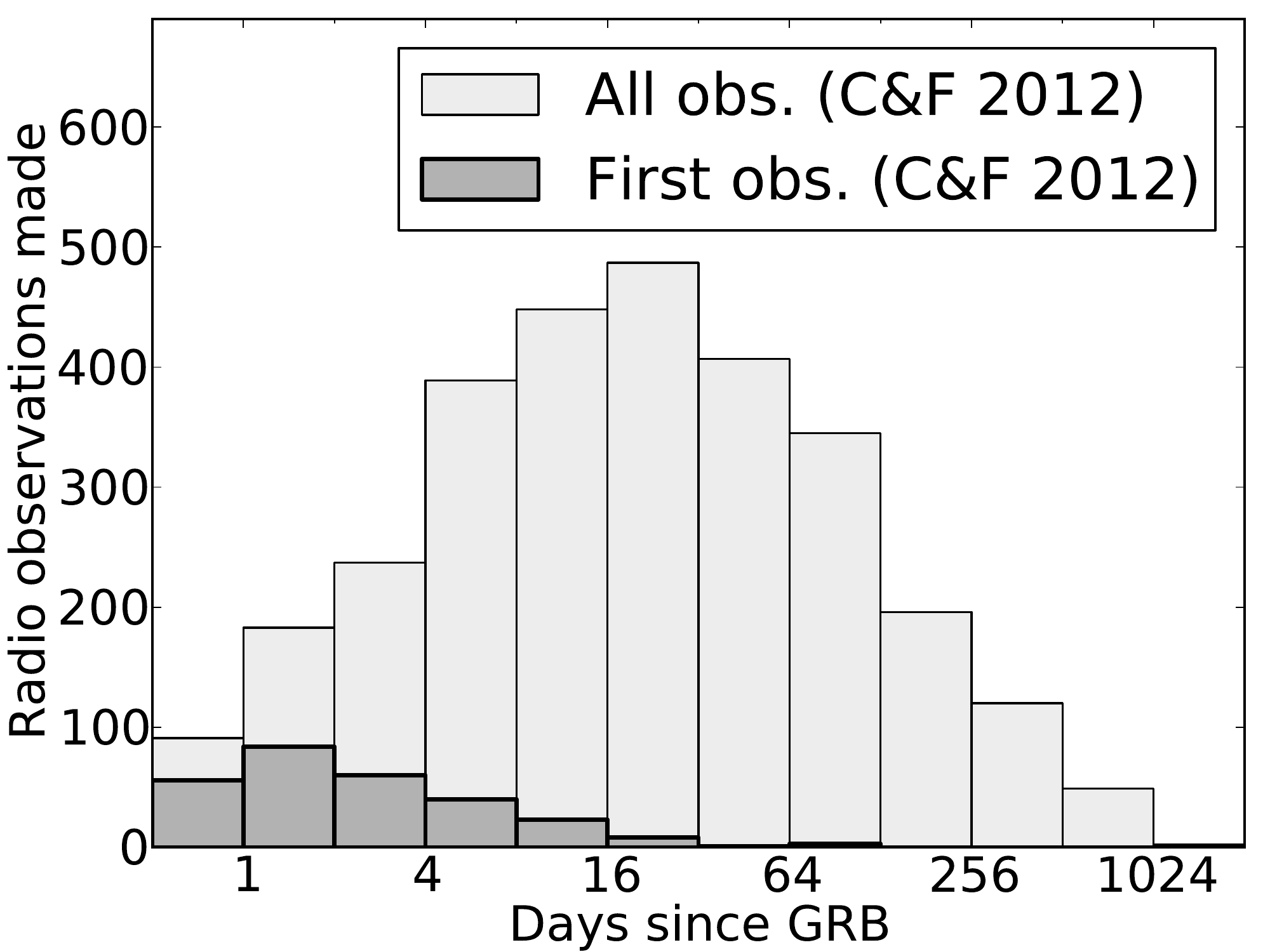}
    \includegraphics[width=0.48\textwidth]{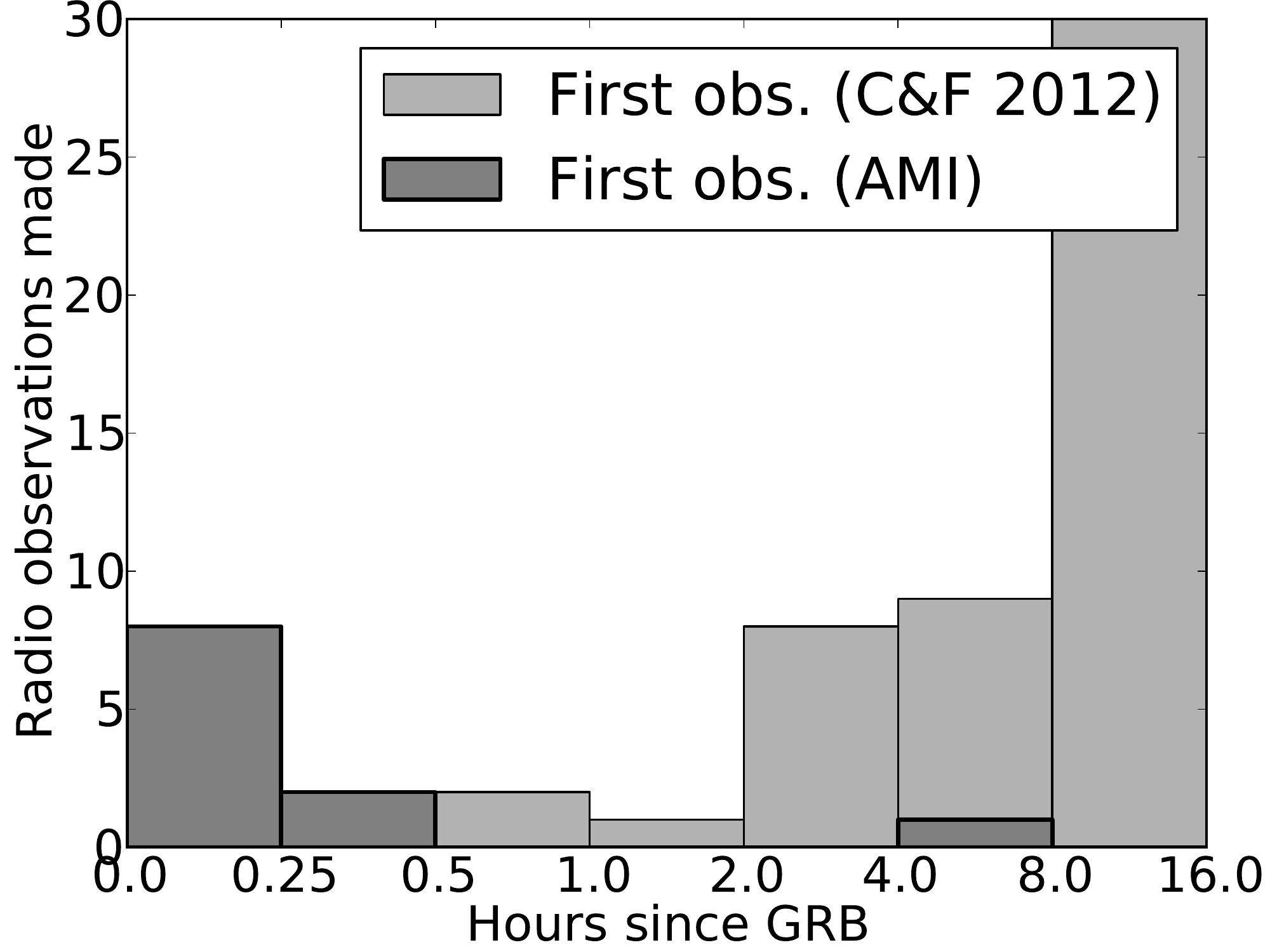}
  
\caption[GRB radio follow-up delay histograms]{%
GRB radio follow-up delay histograms. 
  Top: A histogram depicting radio observations of GRBs made with the VLA. 
  Light grey
  bars show counts for all observations, while 
  medium grey 
  bars (overlaid) show counts for the first observation of any given source. 
  Bottom: A close up depicting the observation counts for initial radio observations of a GRB source undertaken less than 16 hours after the burst. 
  Medium grey
  bars depict VLA observations. 
  Dark grey
  bars, overlaid, depict observations presented in this paper (there are no catalogued VLA observations within the 30 minute period immediately after a burst). Data on VLA observations is reproduced from \cite{Chandra2012}.
\label{fig:cf2012}
} 
\end{center} 
\end{figure}

Immediate observations with AMI are not always possible --- the telescope may be undergoing maintenance, experiencing unfavourable weather conditions, or the source may simply be over the horizon. 
However, even initial observations delayed by a few hours are faster than the large majority of previously catalogued radio follow-up, as illustrated in Figure~\ref{fig:cf2012}.

\subsection{Radio follow-up lightcurves}
\label{sec:lightcurves}
\begin{figure*}
\begin{center}  
  \includegraphics[width=0.48\textwidth]{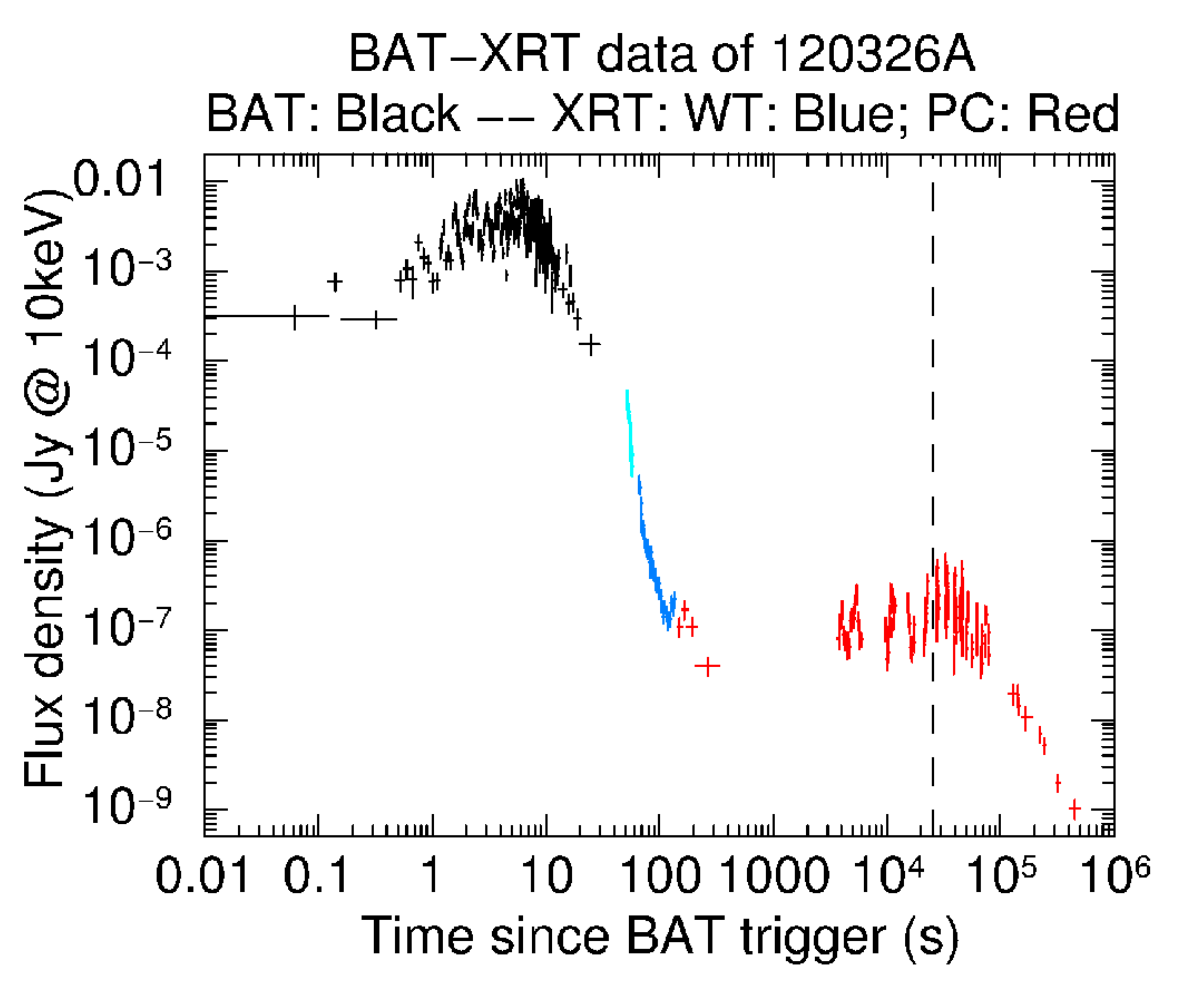}
  \includegraphics[width=0.5\textwidth]{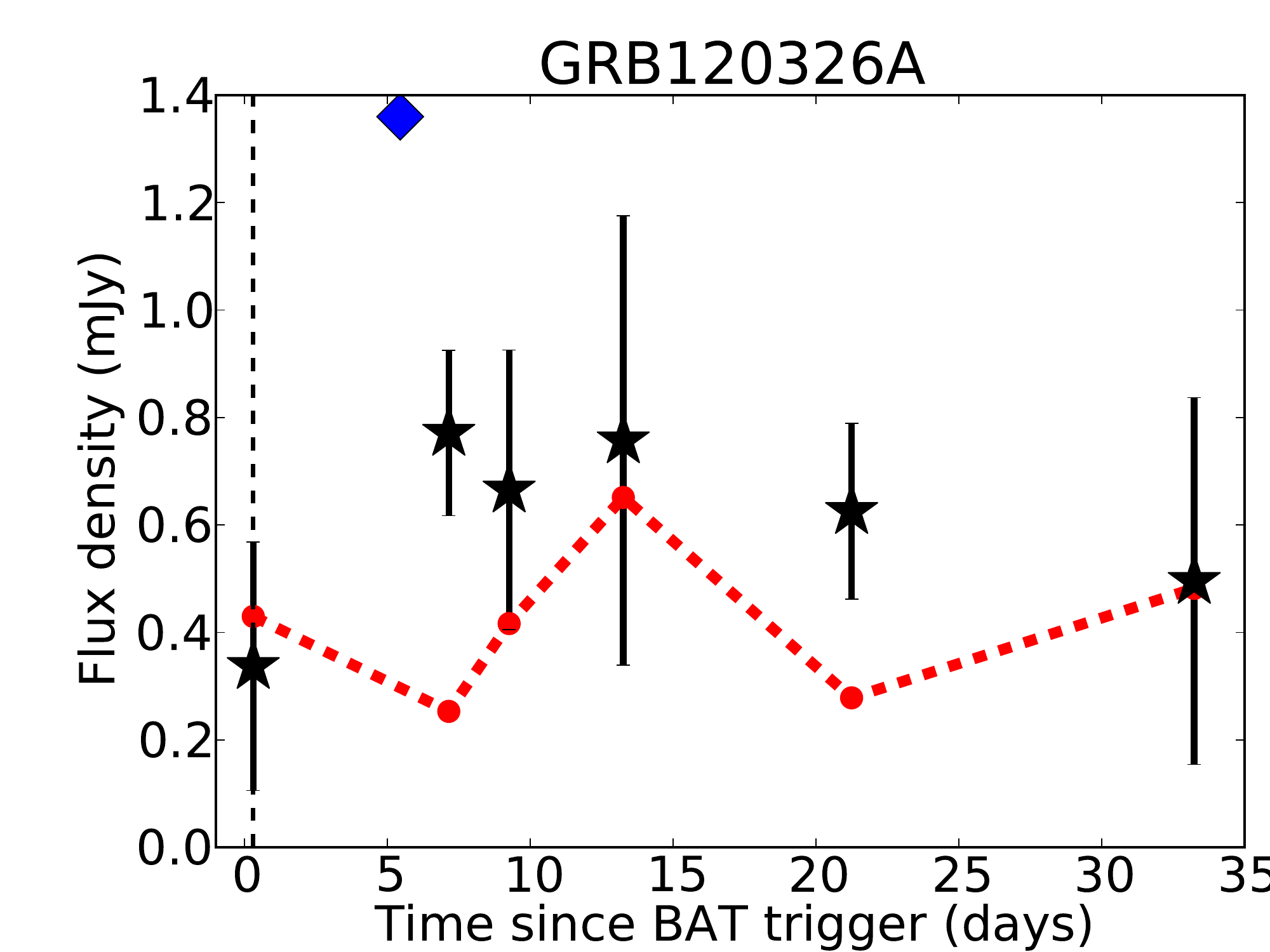}
 \caption[GRB120326A lightcurves]{%
Lightcurves for GRB120326A. 
Left: \textit{Swift} BAT and XRT measurements of the gamma- and X-ray emission. 
Right: Radio-afterglow lightcurve. 
Star points with errorbars represent 15-GHz observations made with AMI as part of this programme. 
Circle points represent $3\sigma$ blind detection thresholds --- a connecting dashed line is overplotted to guide the eye. 
The 
diamond point at 5.45 days represents a JVLA detection at 22\,GHz reported via GCN circulars \citep{Laskar2012}. 
Vertical dashed line in both plots represents the time at which AMI first started taking data on the GRB source position. 
Note that GRB 120326A was observed after a much longer delay than is possible with the automated response system ($\sim$7 hours rather than the typical 5 minutes).
\label{fig:lightcurves}
} 
\end{center} 
\end{figure*}

Delays between each GRB trigger and our initial follow-up observation, and corresponding $3\sigma$ upper limits on the flux density, are given in Table~\ref{tab:init_obs}.
We detect a radio afterglow with confidence level above $5\sigma$ for GRB 120326A, as shown in Figure~\ref{fig:lightcurves}. 
The remaining ten have lightcurves consistent with non-detection. 
Specifically, in this preliminary analysis we do not detect any radio emission from the five GRBs observed within the first six hours post-GRB trigger. This places upper limits of less than $1\milliJans$ on the $15$-GHz flux density averaged over the one hour observation period, but only weakly constrains possible emissions on much shorter timescales. More detailed analysis of the data and upper limits is planned for future work, as discussed in Section~\ref{sec:discuss}.  
A full listing of all observations  is given in Appendix~\ref{app:obs_table}. 
The BAT and BAT-XRT lightcurves shown in Figure~\ref{fig:lightcurves} were obtained from the automated burst analyser page at the UK {\it Swift} Science Data Centre Website \citep{evans2010} using $4\,\sigma$ significance bins and plotting the observed flux density at 10 keV. 

\section{Discussion and future work}
\label{sec:discuss}
In summary, this paper presents the first results from our programme of GRB follow-up observations using AMI-LA, demonstrating the success of the fully automated rapid-response follow-up system we have developed. 
Future observations will provide the first large sample of early-time radio observations of GRBs, placing tight constraints on theoretical models, while our monitoring strategy will result in a sample of radio afterglows unbiased by selection effects.

There are a number of enhancements to the AMI-LA response system we hope to implement in the near future. These include:
\begin{itemize}
 \item Notifying the wider astronomical community when AMI-LA follow-up observations have been made, both through the NASA-GCN and the VOEvent network.
 \item Continued refinement of the data reduction process, including automated data transfer and initiation of the data reduction process. 
    We will also implement time-slicing and analysis of the data on timescales shorter than one hour, to look for short-duration, high intensity emission. 
    This is a particularly exciting avenue of investigation given recent results reported in \cite{Bannister2012}.
 \item Automated discovery and user notification of GRB afterglow candidates, to enable rapid human intervention and detailed follow-up.
\end{itemize}
We note that a human-verified dataset of modest size such as this provides an excellent test bed for automated transient-detection and classification tools. We are working closely with the LOFAR Transients Key Project team on software to this end \citep[][]{Fender2006,Swinbank2011}, and intend to employ it to automate much of the data curation currently done manually in the course of these follow-up observations.

This work was made possible in part through ERC funding for the 4 Pi Sky project.\footnote{http://www.4pisky.soton.ac.uk} 
In a wider context, this work provides a first step toward the longer term goals of that project, such as automated data reduction and distribution of \emph{machine readable} data products (not just notifications) to the astronomical transients community, and use of such data products to enable rapid, automated decisions on allocation of telescope resources for transient follow-up.

It is hoped that the software described in this paper will be useful to the wider astronomical transients community in working toward these aims. We also plan to implement follow-up protocols in collaboration with other observatories. 
Notably, we are collaborating with the LOFAR-UK station \citep{Fender2010} which will provide extremely rapid response due to the way in which LOFAR stations are made up of phased arrays (and hence have no moving parts). We plan to start taking data with LOFAR-UK in the near future.

\section*{Acknowledgements}
This work makes use of GRB alerts supplied via NASA GCN, and of data supplied by the UK \textit{Swift} Science Data Centre at the University of Leicester. 
We thank the staff of the Mullard Radio Astronomy Observatory for
their assistance in the commissioning and operation of
AMI, which is supported by Cambridge University and the STFC. 
We also thank Poonam Chandra for supplying the delay times of observations in the \cite{Chandra2012} catalogue. 
This project was funded in part by European Research Council Advanced Grant 267697 ``4 pi sky: Extreme Astrophysics with Revolutionary Radio Telescopes.''

\bibliographystyle{mn2e}
\bibliography{radio_astronomy_and_GRB_refs_MNRAS_abbrev}

\appendix
\section{Table of observations}
A full list of all AMI-LA 15-GHz observations presented in this paper is given in table~\ref{tab:obs_all}.
\label{app:obs_table}
\begin{table*}
\begin{minipage}{0.45\linewidth}
%
%
%
%
\begin{tabular}{ l  r  r  r }
\toprule
 \parbox{ 0.1 \textwidth}{ \begin{center}\textbf{ Start date (UTC) } \end{center} } 
 &   \parbox{ 0.1 \textwidth}{ \begin{center}\textbf{ Days since burst } \end{center} } 
 &   \parbox{ 0.1 \textwidth}{ \begin{center}\textbf{ Flux density (mJy) } \end{center} } 
 &   \parbox{ 0.15 \textwidth}{ \begin{center}\textbf{ Image std. dev. (mJy) } \end{center} } 
  \\
\toprule

    \multicolumn{4}{c}{ \textbf{ GRB120305A }} \\

            12/03/05 19:41:57 & \textless0.01 & \textless 0.316  &  0.105 \\
        
            12/03/06 14:20:54 & 0.78 & \textless 0.158  &  0.053 \\
        
            12/03/07 15:26:46 & 1.83 & \textless 0.268  &  0.089 \\
        
            12/03/08 15:52:45 & 2.84 & \textless 0.171  &  0.057 \\
        
            12/03/09 18:08:27 & 3.94 & \textless 0.227  &  0.076 \\
        
            12/03/10 16:14:49 & 4.86 & \textless 0.193  &  0.064 \\
        
            12/03/11 17:50:36 & 5.93 & \textless 0.293  &  0.098 \\
        
            12/03/12 14:57:08 & 6.81 & \textless 0.272  &  0.091 \\
        
            12/03/13 17:17:49 & 7.90 & \textless 0.218  &  0.073 \\
        
            12/03/15 17:44:51 & 9.92 & \textless 0.226  &  0.075 \\
        
            12/03/16 18:00:52 & 10.93 & \textless 0.163  &  0.054 \\
        
            12/03/19 17:29:07 & 13.91 & \textless 0.203  &  0.068 \\
        
            12/03/22 13:08:01 & 16.73 & \textless 0.221  &  0.074 \\
        
            12/03/27 16:17:47 & 21.86 & \textless 0.197  &  0.066 \\
        
            12/04/01 16:18:04 & 26.86 & \textless 0.194  &  0.065 \\
        
            12/04/05 16:42:14 & 30.88 & \textless 0.169  &  0.056 \\

     \midrule 

    \multicolumn{4}{c}{ \textbf{ GRB120308A }} \\

            12/03/08 06:18:20 & \textless0.01 & \textless 0.164  &  0.055 \\
        
            12/03/09 02:16:03 & 0.83 & \textless 0.144  &  0.048 \\
        
            12/03/13 03:20:06 & 4.88 & \textless 0.267  &  0.089 \\
        
            12/03/16 06:02:50 & 7.99 & \textless 0.183  &  0.061 \\
        
            12/03/19 05:11:09 & 10.96 & \textless 0.199  &  0.066 \\
        
            12/03/22 03:54:32 & 13.90 & \textless 0.217  &  0.072 \\
        
            12/03/25 03:27:46 & 16.88 & \textless 0.210  &  0.070 \\
        
            12/04/02 03:46:11 & 24.90 & \textless 0.169  &  0.056 \\
        
            12/04/11 02:25:55 & 33.84 & \textless 0.190  &  0.063 \\

     \midrule 

    \multicolumn{4}{c}{ \textbf{ GRB120311A }} \\

            12/03/11 05:37:37 & \textless0.01 & \textless 0.230  &  0.077 \\
        
            12/04/05 04:39:13 & 24.96 & \textless 0.186  &  0.062 \\
        
            12/04/08 03:37:33 & 27.92 & \textless 0.180  &  0.060 \\

     \midrule 

    \multicolumn{4}{c}{ \textbf{ GRB120320A }} \\

            12/03/21 02:15:44 & 0.60 & \textless 0.196  &  0.065 \\
        
            12/04/05 01:24:45 & 15.56 & \textless 0.212  &  0.071 \\

     \midrule 

    \multicolumn{4}{c}{ \textbf{ GRB120324A }} \\

            12/03/24 06:03:17 & \textless0.01 & \textless 0.218  &  0.073 \\
        
            12/04/05 05:49:01 & 11.99 & \textless 0.175  &  0.058 \\
        
            12/04/08 04:47:22 & 14.95 & \textless 0.240  &  0.080 \\
        
            12/04/16 08:25:13 & 23.10 & \textless 0.184  &  0.061 \\
        
            12/04/23 03:58:21 & 29.92 & \textless 0.329  &  0.110 \\

\bottomrule
\end{tabular}

\end{minipage}
\hfil
\begin{minipage}{0.45\linewidth}
%
%
%
%
\begin{tabular}{ l  r  r  r }
\toprule
 \parbox{ 0.1 \textwidth}{ \begin{center}\textbf{ Start date (UTC) } \end{center} } 
 &   \parbox{ 0.1 \textwidth}{ \begin{center}\textbf{ Days since burst } \end{center} } 
 &   \parbox{ 0.1 \textwidth}{ \begin{center}\textbf{ Flux density (mJy) } \end{center} } 
 &   \parbox{ 0.15 \textwidth}{ \begin{center}\textbf{ Image std. dev. (mJy) } \end{center} } 
  \\
\toprule

    \multicolumn{4}{c}{ \textbf{ GRB120326A }} \\

            12/03/26 08:42:58 & 0.31 & 0.337  &  0.143 \\
        
            12/04/02 04:55:59 & 7.15 & 0.771  &  0.084 \\
        
            12/04/04 07:37:40 & 9.26 & 0.666  &  0.139 \\
        
            12/04/08 07:31:55 & 13.26 & 0.757  &  0.217 \\
        
            12/04/16 07:15:25 & 21.25 & 0.626  &  0.093 \\
        
            12/04/28 07:03:08 & 33.24 & 0.495  &  0.161 \\

     \midrule 

    \multicolumn{4}{c}{ \textbf{ GRB120403A }} \\

            12/04/03 08:34:27 & 0.31 & \textless 0.238  &  0.079 \\
        
            12/04/04 16:06:16 & 1.63 & \textless 0.205  &  0.068 \\
        
            12/04/06 16:08:23 & 3.63 & \textless 0.204  &  0.068 \\
        
            12/04/10 14:52:49 & 7.57 & \textless 0.212  &  0.071 \\
        
            12/04/16 15:34:03 & 13.60 & \textless 0.254  &  0.085 \\
        
            12/05/01 14:50:02 & 28.57 & \textless 0.349  &  0.116 \\

     \midrule 

    \multicolumn{4}{c}{ \textbf{ GRB120404A }} \\

            12/04/05 00:09:57 & 0.47 & \textless 0.223  &  0.074 \\
        
            12/04/08 00:18:06 & 3.48 & \textless 0.237  &  0.079 \\
        
            12/04/11 00:11:17 & 6.47 & \textless 0.169  &  0.056 \\
        
            12/04/17 01:57:21 & 12.55 & \textless 0.214  &  0.071 \\

     \midrule 

    \multicolumn{4}{c}{ \textbf{ GRB120422A }} \\

            12/04/22 13:36:43 & 0.27 & \textless 0.616  &  0.205 \\
        
            12/04/24 16:17:23 & 2.38 & \textless 0.465  &  0.155 \\
        
            12/04/26 16:19:30 & 4.38 & \textless 0.368  &  0.123 \\
        
            12/04/29 17:17:31 & 7.42 & \textless 0.239  &  0.080 \\
        
            12/05/05 17:53:45 & 13.45 & \textless 0.228  &  0.076 \\
        
            12/05/20 21:19:03 & 28.59 & \textless 0.456  &  0.152 \\

     \midrule 

    \multicolumn{4}{c}{ \textbf{ GRB120514A }} \\

            12/05/16 04:07:39 & 2.12 & \textless 0.211  &  0.070 \\
        
            12/05/19 03:01:00 & 5.08 & \textless 0.460  &  0.153 \\
        
            12/05/22 01:54:22 & 8.03 & \textless 0.279  &  0.093 \\
        
            12/06/03 23:58:25 & 20.95 & \textless 0.619  &  0.206 \\
        
            12/06/11 03:20:16 & 28.09 & \textless 1.488  &  0.496 \\

     \midrule 

    \multicolumn{4}{c}{ \textbf{ GRB120521C }} \\

            12/05/21 23:36:44 & 0.01 & \textless 0.302  &  0.101 \\
        
            12/05/23 21:02:16 & 1.90 & \textless 0.793  &  0.264 \\
        
            12/05/25 20:54:25 & 3.90 & \textless 0.307  &  0.102 \\
        
            12/05/28 21:02:34 & 6.90 & \textless 0.328  &  0.109 \\
        
            12/06/04 18:30:23 & 13.80 & \textless 0.216  &  0.072 \\
        
            12/06/23 18:50:25 & 32.81 & \textless 0.342  &  0.114 \\

\bottomrule
\end{tabular}

\end{minipage}

\caption{
15-GHz measurements of the first eleven GRB sources observed with AMI-LA. 
For sources where radio emission was not detected (i.e. all except GRB 120326A), the flux density is given as an upper limit at the $3\sigma$ level.
\label{tab:obs_all}
}
\end{table*}

\label{lastpage}
\end{document}